\documentclass[12pt,a4paper]{article}
\usepackage[T2A]{fontenc}
\usepackage[cp1251]{inputenc}
\usepackage[english]{babel}
\usepackage{amssymb}
\usepackage{amsmath}
\usepackage{graphicx}
\usepackage[small,sc,center]{titlesec}
\usepackage{indentfirst}
\usepackage{caption2}
\newcommand{\msun}{\,$M_{\odot}$}

\newcommand{\kms}{\,km\,s$^{-1}$}

\begin{document}
	
\begin{center}
\textbf{\large Cosmic abundance of iron} 

\vskip 5mm
\copyright\quad
2023 г. \quad N. N. Chugai\footnote{email: nchugai@inasan.ru}\\

\textit{$^1$Institute of Astronomy, Russian Academy of Sciences, Moscow} \\
\vspace{0.5cm}

Submitted  15.05.2023 г.
\end{center}


\noindent
{\em PACS codes:\/}

\clearpage
 
 \begin{abstract} 
I explore a possibility to estimate an upper limit of the current iron abundance of the barion matter. 
The upper limit is determined by the minimal iron abundance, at which the gamma-ray background, produced by 
the decay of $^{56}$Ni synthesised in the Universe to date, contradicts the observational MeV gamma-ray background. 
I calculate the gamma-ray background from  SNe~Ia and SNe~II with the gamma-ray scattering and absorption in 
supernova envelope. It is shown that the model background does not contradict the observed MeV background, 
if the present day iron abundance of the barion matter is less than 
15\% of the solar abundance.

\end{abstract}

\section{Introduction}

Overwhelming fraction of iron (92\%) is represented by the isotop 
$^{56}$Fe that is synthesised by supernovae as $^{56}$Ni. It converts 
into iron via two step decay $^{56}$Ni (8.8\,d) -- $^{56}$Co 
(111.26\,d) -- $^{56}$Fe (Nadyozhin 1994).
The  $^{56}$Ni synthesis by SNe~II (SN~1987A) is demonstrated via the detection of gamma-ray lines from the $^{56}$Co decay by the {\em SMM} observatory (Matz et al. 1988) and hard X-ray radiation detected by the orbital observatory {\em Kvant} (Sunyaev et al. 1987).  In the case of SN~Ia (SN~2014J) the $^{56}$Ni synthesis is demonstrated via the detection of gamma-ray 
lines from the $^{56}$Co decay by the {\em INTEGRAL} observatory 
(Churazov et al. 2014.)

Based on the fact that all the iron in the Universe originates from 
$^{56}$Ni decay ejected by supernovae Clayton and Silk (1969) estimated 
the brightness of the diffuse cosmic background produced by gamma-quanta of radioactive decay as $3.3\times10^{-2}$ cm$^{-2}$\,s$^{-1}$\,sr$^{-1}$,
that turned out comparable to the observed MeV-band background. 
Later, Clayton and Ward (1975) supported this con\-clusion based on the 
comparison with the background obtained at {\em Appollo 15}.
Of course, for the background calculations authors used the barion density and Hubble constant significantly different compared to 
nowdays values; moreover the adopted solar abundance is rather  
unrealistic assumption. 
 
The latter remarks motivate one to revisit this kind of analysis and 
pose somewhat different question: whether the measurement of the 
MeV background could be used to estimate an upper limit of the present day iron abundance of the barion matter?
The question is intriguing since the answer is unknown, whereas it 
could clarify a general picture of the star formation and nucleosynthesis in the Universe.
It should be emphasised that the posed question differs from the task of 
the gamma-ray background computation based on available estimates of 
supernova rates (Ruiz-Lapuente et al 2001, 2016; Iwabuchi and Kumagai 2001; Horiuchi et al. 2010; Lacki et al. 2014).
 
An attempt to answer the question on the upper limit of the iron abundance in the barion matter is the primary goal of this paper.
Generally, the problem is reduced to the calculation of the gamma-ray 
background produced by the $^{56}$Ni decay for the adopted present day 
$^{56}$Fe abundance in the barion matter and adopted dependence of 
normalized rate of supernovae on the redshift.
A major difference with earlier works (Clayton and Silk 1969, Clayton and Ward 1975), apart from the new data on barion density and Hubble constant, is the account for the gamma-quanta transfer in the expanding shell of SN~Ia and SN~II. 

Preliminary considerations essential for the gamma-ray background \\
computation, particularly, relative contribution  of SNe~Ia and SNe~II into the iron synthesis and redshift dependence of the supernova rates, are considered in the next section.
Hereafter we use cosmological parameters $\Omega_m = 0.3$, $\Omega_b = 0.046$, $\Omega_{\Lambda} = 0.7$ and $H_0 = 70$\kms\,Mpc$^{-1}$.

\section{Supernovae and iron synthesis}

\subsection{Relative role of SN~Ia и SN~II}

The first estimate of the relative role of different supernovae in the iron synthesis suggested that almost all the galactic iron could be produced by SNe~II (Arnett et al. 1989).
Later, Thielemann et al. (2002) have concluded that SNe~Ia contribute 
50-60\% to the present day iron. 
This estimate was obtained from the observed ratio SN~Ia/SN~II of 
extragakactic supernovae assuming iron production 
$m_1(\mbox{Fe}) = 0.6$\msun\ and $m_2(\mbox{Fe}) = 0.1$\msun\ per one SN~I\label{key}a and SN~II, respectively.
 
One can use an alternative approach to estimate the relative role of supernovae based on the evolution of stellar iron and oxygen abundance.
The O/Fe ratio in low metallicity stars, viz., [Fe/H] $\equiv \lg\,[\mbox{(Fe/H)/(Fe/H)}_{\odot}$] $\sim$ -2.6... -2, demonstrates a plateau on the 
[O/Fe] vs. [Fe/H] diagram at the level of [O/Fe] $\approx 0.7$ (Sitnova \& Mashonkina 2018).
Since the initial galactic nucleosynthesis is dominated by core-collapse supernovae, the inferred [O/Fe] value for old stars 
indicates that the average O/Fe ratio per one SN~II exceeds the solar ratio by a factor of $A$ = (O/Fe)/(O/Fe)$_{\odot} \approx 5$.
 
The solar ratio O/Fe can be expressed via the total mass of synthesised oxygen and iron by supernovae SN~Ia and SN~II		
($M_1$ и $M_2$). Neglecting the SN~Ia contribution to the galactic oxygen, 
one can write down the solar ratio O/Fe as 
\begin{equation}   
\left(\frac{O}{Fe}\right)_{\odot} =  \frac{M_1(\mbox{O}) + 
	M_2(\mbox{O})}{M_1(\mbox{Fe}) + M_2(\mbox{Fe})} \approx A\left(\frac{O}{Fe}\right)_{\odot}(1 + \mu_{12})^{-1}\,,
\label{eq:ofe}
\end{equation}
where $\mu_{12} = M_1(\mbox{Fe})/M_2(\mbox{Fe})$  is the relative contribution of SN~Ia/SN~II to the iron synthesis. 
This relation immediately gives us desirable value $\mu_{12} \approx 4$, which 
means that SNe~Ia provide 80\% of synthesised iron, whereas 20\% of iron come  from SNe~II.

%================================================================
\begin{figure}
	\centering
	\includegraphics[width=0.97\textwidth,clip]{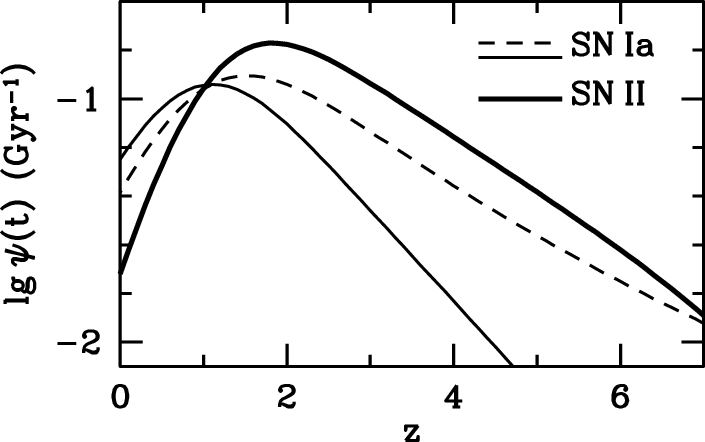}
	\caption{
		Comoving normalized supernova rate versus redshift. 
		The rate of SN~II ({\em thick line}) is proportional to SFR, whereas 
		SN~Ia rate is the SFR convolved with the DTD for $\tau_0 = 10^8$\,yr ({\em dashed line}) and  $\tau_0 = 5\times10^8$\,yr ({\em thin solid line}).
	}
	\label{fig:sfr}
\end{figure}
%==================================================================

\subsection{Evolution of supernova rates}

To calculate the gamma-ray background one needs to know the normalized 
dependence of supernova rates on the redshift.
The rate of SNe~II can be adopted to be proportional to the star formation rate (SFR), since the delay between progenitor birth and explosion for $M > 9$\msun\ is small compared to the SFR time scale (cf. Madau \&  Dickinson 2014). 
The SFR is presented here (Figure 1) by the broken power law 
$\psi_0 \propto (t/t_{br})^q$ with the brake at $t_{br} = 3.6\times10^9$ yr and power $q = 1.2$ and -3.4 for $t < t_{br}$ and $t > t_{br}$ respectively. 

The SN~Ia rate is related to the SFR in a complicated way.
In a paradigm of binary scenario (double white dwarf or white dwarf with a non-degenerate star) the evolution towards the explosion can take time comparable to the age of Universe (Tutukov \& Yungelson 1994). 
In theory, the SN~Ia rate is described by the convolution of the SFR and the 
 delay time distribution (DTD) function  $p(\tau)$, with $\tau$ being the delay of the explosion wrt binary birth. 
Ones consider sometimes nonmonotonic functions $p(\tau)$ in an attempt to take into account different evolution scenaria and explosion models (Kobayashi et al. 2020), 
however I adopt power law $p(\tau) \propto \tau^{-1}$ (Frohmaier et al. 2019).
The function $p(\tau)$ is non-zero for $\tau > \tau_0$ and the minimal delay 
$t_0$ lies between $10^8$\,yr (Tutukov \& Yungelson 1994) and $5\times10^8$\,yr (Yubgelson 2010; Kobayashi et al. 2020). 
I adopt the explosion scenario of Chandrasekhar CO white dwarf that 
most adequitely reproduces the [O/Fe] vs. [Fe/H] diagram (Kobayashi et al. 2020) and consider two cases $\tau_0 = 10^8$\,yr and $\tau_0 = 5\times10^8$\,yr.
 
Both options of the SN~Ia rate, as well as the SN~II rate in the 
normalised version (integral equals unity) are shown in Figure 1.
The age -- redshift relation is determined as $dt = da/\dot{a}$, where 
the dimensionless Universe radius is $a = 1/(1 + z)$ and the expansion rate 
neglecting radiation is   
$\dot{a}/a = H_0[\Omega_m(1 + z)^3 + \Omega_{\Lambda}]^{1/2}$ (Peebles 1993). 

\section{Gamma-ray background from $^{56}$Ni decay}

\subsection{Model}

The photon brightness of the gamma-ray background from sources with the isotropic distribution along the redshift and the emissivity 
$4\pi j = g(\epsilon,z)$ (cm$^{-3}$\,s$^{-1}$\,MeV$^{-1}$) can be computed 
in a twofold way.
In the first approach ones compute the gamma-ray density via integration of 
the source density over volume (e.g. Ruiz-Lapuente et al. 2016). 
Alternatively, one can directly calculate the background photon brightness 
$\phi(\epsilon_0)$ (cm$^{-2}$\,s$^{-1}$\,MeV$^{-1}$\,sr$^{-1}$) as a 
formal solution of the radiation transfer with isotropic sources and without absorption 
along the ray
 \begin{equation}
\phi(\epsilon_0) = \frac{c}{4\pi}\int_{0}^{t_{obs}}\frac{g(\epsilon,z)dt}
{(1+z)^2}\,,
\label{eq:int1}
 \end{equation}  
where the photon energy $\epsilon$ emitted at the redshift $z$ is related to 
the observed energy $\epsilon_0$ as $\epsilon = (1+z)\epsilon_0$. 
The factor $1/(1+z)^2$ takes into account the reduction of both the energy 
interval and the photon arrival rate in the observer frame.
This approach is applied, e.g., by Peebles (1993) to calculate the optical background. 
Using the differential relation $dt/dz =1/[H_0(1+z)E(z)]$, where 
$E(z) = [(1+z)^3\Omega_m + \Omega_{\Lambda}]^{1/2}$, the integration over time is reduced to the intergration over $z$ 
 \begin{equation}
\phi(\epsilon_0) = \frac{c}{4\pi H_0}\int_{0}^{z_{max}}\frac{g(\epsilon,z)dz}{(1+z)^3E(z)}\,.
\label{eq:int2}
\end{equation}  
Background brightness is the sum of contribution of SNe~Ia и SNe~II 
$\phi = \phi_1 + \phi_2$, where  $k = 1$ and 2 for SN~Ia and SN~II, respectively. 
Emissivity for each supernova type is    
 \begin{equation}
g_k(\epsilon,z) = \omega_k f(56)\rho_0(1 + z)^3\Omega_bX\Psi_k(z)(N_A/56)\Phi_k(\epsilon)\,,
\end{equation}  
where $\omega_1 = 0.8$ and $\omega_2 = 0.2$, $f(56) = 0.92$ is the $^{56}$Fe isotope fraction, $\rho_0$ is the 
critical density, $\Omega_b$ is the barion fraction, $X$ is the present day iron abundance in the barion matter, $\Psi_k(z)$ is the normalized rate of the $^{56}$Ni production for a certain supernova type, $N_A$ is the Avogadro number, $\Phi_k(\epsilon)$ (MeV$^{-1}$) is the spectrum 
of escaping gamma-quanta  per one $^{56}$Ni nucleus integrated over 600 days  with the scattering and absorption in the envelope taken into account.

At first glance there might be significant uncertaity related to the absence of 
a consensus on the preferred explosion model for SNe~Ia.
Yet computations (Ruiz-Lapuente et al. 2016) of three different models 
including fully mixed W7 (Nomoto et al. 1984), 
3D-model with the delayed detonation (R\"{o}pke et al. 2012), and binary CO white dwarf merging with the subsequent explosion (Pakmor et al. 2012) 
result in the identical gamma-ray spectra in the energy band of the maximum flux.
This fact is related to the similar model ejecta mass and energy along with the 
 almost complete mixing of $^{56}$Ni, which suggests 
similar column density at the same age.

Above mentioned gives us a reason to consider relatively simple SN~Ia model:
a freely expanding envelope of the  uniform density and composition. 
The adopted below ejecta mass of 1.4\msun\ and the kinetic energy 
of $1.3\times10^{51}$\,erg (or 1.3\,B) correspond to the W7 model (Nomoto et al. 1984). 
For the "average" SN~II we adopt ejecta mass of 13\msun\ that corresponds to the initial progenitor mass of 15\msun,  kinetic energy of $10^{51}$ \,erg, and assume $^{56}$Ni mixing in the central zone of 2.8\msun;
the variation of the latter value does not  
 affect the total gamma-ray spectrum in the energy range of 
the maximum flux.

%================================================================
\begin{figure}
	\centering
	\includegraphics[width=0.97\textwidth]{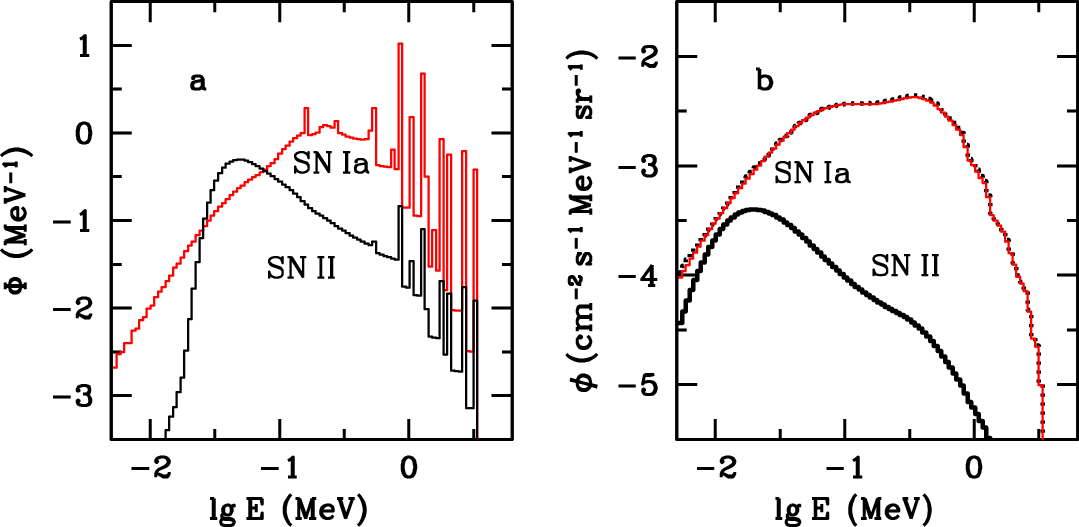}
	\caption{	
		{\em Left.} Rest frame gamma-ray spectra of SN~Ia ({\em red}) and SN~II 
		per one $^{56}$Ni nucleus and accumulated during 600 days after explosion.
		{\em Right.} Brightness of the gamma-ray diffuse background from SN~Ia 
		with the explosion energy of 1.3\,B ({\em dotted line}) and 1\,B ({\em thin solid line}), 
		and from SN~II with 80\%  of iron from SNe~Ia and 
		20\% from SNe~II assuming present day iron abundance in the barion matter 
		$X(\mbox{Fe}) = 0.15X(\mbox{Fe})_{\odot}$. 
	}
	\label{fig:fig2}
\end{figure}
%==================================================================

\subsection{Modelling gamma-ray background}

The spectrum of escaping gamma-rays from the  $^{56}$Ni -- $^{56}$Co -- $^{56}$Fe decay in the rest frame is calculated using the Monte Carlo technique. 
The separate spectra computed every 6 days through 600 days are 
stacked into the resulting spectrum. 
Emitted photon energy and probability per one decay are taken from 
(Nadyozhin 1994).  

Generally, emitted photon experiences multiple Compton scattering with the Klein-Nishina cross section and can eventually either escape the envelope or 
be absorbed. 
Absorption coefficients are taken from the NIST database. 
The SN~Ia composition is represented by equal fractions of Fe and Si.
The difference in the absorption coefficient between Fe, Ni, and Co is 
small and permits us to neglect the composition change due to the decay.
For SN~II we take into account absorption by O, Mg, Si, Fe with the 
iron mass of 0.044\msun\ in line with average $^{56}$Ni mass for SNe~II 
(Anderson 2019) with masses of O, Mg, and Si being equal equal 0.68, 0.027 and 0.11\msun, respectively, in line with the model s15A of Woosley and Weaver (1995).
The indicated masses are combined with masses of these elements in the presupernova 
 hydrogen envelope assuming solar abundance. 
 
The rest-frame gamma-ray spectra of SN~Ia and SN~II per one $^{56}$Ni nucleus 
are shown in Figure 2a.
The spectra are qualitatively consistent with those calculated 
earlier (Watanabe et al. 1999, Iwabuchi \& Kumagai 2001). 
The calculated cosmic background produced by SN~Ia (80\%) and 
SN~II (20\%) assuming the present day iron abundance in the barion matter $X(\mbox{Fe}) = 0.15X(\mbox{Fe})_{\odot}$ are shown in Figure 2b.
In the case of SN~Ia two versions, with the  explosion energy of 1.3\,B and 1.1\,B, are presented. 
For the fixed mass the energy $E$ determines the age when the envelope becomes transparent for MeV gamma-quanta, $t \propto 1/\sqrt{E}$.   
The Figure 2b demonstrates weak dependence on the energy and therefore, on the 
choice of the dominant SN~Ia model.
We use the $^{56}$Ni production rate vs. redshift shown in Figure 1 and the rate of SN~Ia corresponding to $\tau_0 = 10^8$ yr. 
The contribution of supernovae SN~II in the background is rather small 
compared to SN~Ia, which is related to the low  ${56}$Ni production and 
low fraction of escaping quanta.

The calculated gamma-ray background produced by SN~Ia и SN~II for 
the present day cosmic iron abundance $X(\mbox{Fe}) = 0.15X(\mbox{Fe})_{\odot}$ and two cases of time delay of SN~Ia 
explosion, $\tau_0 = 10^8$\,yr and $5\times10^8$\,yr, is displayed in Figure 3 
along with observed MeV diffuse background according to {\em SMM} (Watanabe et al. 1997) and {\em HEAO-1} (Kinzer et al. 1997) data.
The estimate of the supernovae contribution in the observed background is actually 
ill-posed problem in the absence of reliable information on the possible components 
composed the background. 
We therefore reduce this problem to the discussion of the gamma-ray background in the absence of the supernova contribution, symbolically, $bgr_0 = bgr_{obs} - bgr_{sn}$.
The shown average observed  background is the non-linear fit 
with the root mean square deviation of $\sim 2$\%. 

The recovered "observed"\/ background without SNe contribution in the case of $X(\mbox{Fe}) = 0.15X(\mbox{Fe})_{\odot}$ demonstrates pronounced depression 
($\chi^2/\mbox{dof} \approx 10$) in the energy range of background produced 
by supernovae.
The depression position is unnatural and it looks like as a conspiracy 
between different background sources to create a depression just in the energy range where the supernovae contribution is maximal.
This possibility is highly unlikely and should be rejected.  
We therefore conclude that the present day cosmic abundance of iron in the barion matter is less than 15\% of the solar abundance.

%================================================================
\begin{figure}
	\centering
	\includegraphics[width=0.97\columnwidth]{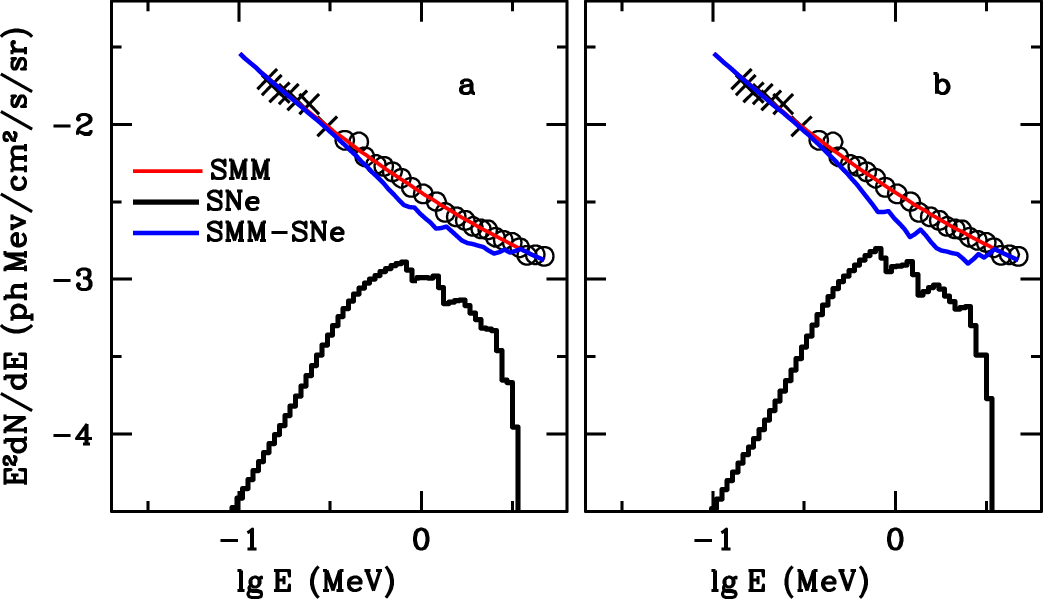}
	\caption{	
		{\em Left.} Gamma-ray background from SN~Ia and SN~II  assuming the present day iron abundance in the barion matter 
		$X(\mbox{Fe}) = 0.15X(\mbox{Fe})_{\odot}$ for $\tau_0 = 10^8$\,yr ({\em thick black| line}) compared to the observed background according to {\em SMM} 
		({\em circles}) and {\em HEAO} ({\em crosses}). {\em Red line} shows non-linear fit of the observed background, whereas {\em blue line} is the difference between the observed background and the calculated background from supernovae.
		{\em Right} The same as left but for $\tau_0 = 5\times10^8$\,yr. 
	}
	\label{fig:fig3}
\end{figure}
%==================================================================

\section{Discussion and conclusions}

The aim of the paper has been to answer a question, what is the upper 
limit of the present-day iron abundance in the barion matter that 
does not contradict to the observational MeV gamma-ray diffuse background. 
The major our result is that the gamma-ray background produced by 
$^{56}$Ni decay is consistent with diffuse MeV background, if 
the present-day cosmic iron abundance is less than 15\% 
of the solar abundance.
Remarkably, the result in not sensitive to the variation of the adopted SN~Ia
explosion energy.

It would be useful to compare this upper limit with the available data on the metallicity of barion matter (Table 1). 
The Table  lists fractions of major barion components $\Omega_i/\Omega_b$ (Nicastro et al. 2018) and their metallicity.
For gas of galaxy clusters and L$\alpha$ forrest the metallicity refers to minimum redshift in available data 
(Balestra et al. 2007, Rafelsky et al. 2012).
For the warm-hot intergalactic medium (WHIM) the fraction of the hot component ($T > 5\times10^5$\,K) is estimated in the range of 9-40\% (Nicastro et al. 2018). 
In the Table  we adopt for this component $\Omega_i/\Omega_b = 0.39$ estimated from the normalazation of all component sum on unity. 
In the bottom line of the Table  we put the average metallicity $\langle Z_b/Z_{\odot}\rangle = 
\Sigma\,(\Omega_i/\Omega_b)(Z/Z_{\odot}) \approx 0.24\pm0.03$.  
 The lower and upper limits correspond to the indicated interval of the WHIM metallicity.

When comparing cosmic iron abundance with the metallicity of major components (L$\alpha$ forrest and WHIM)
one should take into account that the metallicity of latter is estimated from spectra of $\alpha$-elements, 
so the metallicity based on iron [Fe/H] can difer from the metallicity [$\alpha$/H] based on $\alpha$-elements.
Indeed, in our Galaxy for the stellat metallicity [Fe/H] = -0.8, that corresponds to 
(Fe/H) = 0.15(Fe/H)$\odot$, the  overabundance [element/Fe] of 
$\alpha$-particle nuclei C, O, Mg, Si is equal 0.15, 0.4, 0.25, 0.25\,dex, respectivaly (Zhao et al. 2016) 
with the average overabundance [$\alpha$/Fe] $\approx 0.25$.
This rule takes place also for L$\alpha$ forrest absorbers: the average difference of metallicities of L$\alpha$ absorbers in the range of -2.7 < [metals/H]  < -0.2 is 
 [$\alpha$/H] - [Fe/H] $\approx 0.3$\,dex (Rafelsky et al. 2012). 
Above arguments suggest that the iron abundance in the major barion components  
is about 10\% of the solar, in agreement with our upper limit of 15\%.  
%====================================================================
\begin{table}[t]
	\vspace{6mm}
	\centering
	{{\bf Table 1.} Principal components of barion matter}\\
	
		  \bigskip	 

	\begin{tabular}{l|c|c} 
		\hline
		 Component       & $\Omega_i/\Omega_b$  & $Z/Z_{\odot}$  \\
	 	\hline	
	 	&    &       \\	
	 {\bf Galaxies:}  & &	\\	
	 Stars   &    0.07          &    1$^a$ \\
	 Cold gas    &     0.017         &    0.5$^b$ \\
	 Hot gas     &     0.05          &    0.5 \\
	 {\bf Intergalactic gas:}   & &	\\
	 Hot gas of galaxy clusters:  &     0.04         &    0.4$^c$ \\
	 L$\alpha$ forrest    &     0.28         &    0.15$^d$ \\
	 WHIM $T<5\times10^5$K  &     0.15         &  0.1 - 0.2$^e$ \\
	 WHIM $T>5\times10^5$K &     0.39        &   0.1 - 0.2$^e$ \\
			\hline
	\multicolumn{3}{c}  {$\langle Z/Z_{\odot}\rangle = 0.24\pm0.03$} \\
   	\hline
   	\end{tabular}
\medskip

\parbox[]{9cm}{\small $^a$\,Gallazzi et al. (2008), $^b$\,De Cia et al.(2021),  $^c$\,Balestra et al. (2007),	$^d$\,Rafelski et al. (2012), $^e$\,Nicastro et al. (2018)}	
\end{table}
%===================================================================

%\clearpage
\vspace{1cm}
\centerline{\bf References}
\vspace{0.5cm}

\noindent
Anderson J.P.,  Astron. Astrophys. {\bf 628}, A7 (2019)\\
\medskip
Arnett W.D., Schramm D.N., Truran J.W.,  Astrophys. J. {\bf 339}, L25 (1989)\\  
\medskip
Balestra I., Tozzi P., Ettori S. et al.,  Astron. Astrophys. {\bf 462}, 429 (2007)\\ 
\medskip
Churazov E., R. Sunyaev R., Isern J., et al., Nature {\bf 512}, 
406 (2014) \\
\medskip
Clayton D.D., Silk J.,  Astrophys. J. {\bf 158}, L43 (1999)   \\
\medskip
Clayton D.D., Ward R.A., Astrophys. J. {\bf 198}, 241 (1975) \\
\medskip
De Cia A., Jenkins E.B., Fox A.J. et al., Nature {\bf 597}, 206 (2021) \\
\medskip
Frohmaier C., Sullivan M., Nugent P. E. et al., Mon. Not. R. Astr. Soc. {\bf 486}, 2308 (2019)  \\
\medskip
Gallazzi A., Brinchmann J., Charlot S.et al.,
 Mon. Not. R. Astron. Soc. {\bf 383}, 1439 (2008)\\
\medskip
Horiuchi1 S., Beacom J.F.,   Astrophys. J. {\bf 723},  329 (2010)\\ 
\medskip
Iwabuchi K., Kumagai S., Publ. Astron. Soc. Japan {\bf 53}, 669 (2001)\\
\medskip
Kinzer R.L., Jung G.V.,  Gruber D.E. et al., Astrophys. J. {\bf 475}, 361 (1997)\\
\medskip
Kobayashi C., Leung S.-C., Nomoto K. et al., Astrophys. J. {\bf 900}, 179 (2020)\\ 
\medskip
Lacki B.C., Brian C., Horiuchi S., Beacom J.F., Astrophys. J. {\bf 786}, 40  (2014)\\ 
\medskip
Madau P., Dickinson M., Ann. Rev. Astron. Astrophys. {\bf 52}, 415 (2014)\\
\medskip
Matz S. M., Share G. H., Chupp E. L., 
AIP Conference Proceedings, Volume 170, pp. 51-59 (1988).\\
\medskip
Nadyozhin D.K., Astrophys. J. Suppl. {\bf 92}, 527 (1994)\\     
\medskip
Nicastro F., Kaastra J., Krongold Y. et al., Nature {\bf 558}, 406 (2018) \\
\medskip
Nomoto K., Thielemann F.-K. , Yokoi K., Astrophys. J. {\bf 286}, 644 (1984)\\
\medskip
Pakmor R., Kromer M., Taubenberger S. et al.  Astrophys. J. {\bf 747}, L10 (2012) \\
\medskip
Peebles P.J.E., {\em Principles of physical cosmology} (Princeton University Press, Princeton, New Jersey, 1993) \\
\medskip
Rafelski M., Wolfe A.M., Prochaska J.X. et al., Astrophys. J. {\bf 755}, 89 (2012) \\
\medskip
Roepke F.K., Kromer M., Seitenzahl I.R. et al.,  Astrophys. J. {\bf 750}, L19 (2012) \\ 
\medskip
Ruiz-Lapuente1 P., The L.-S., Hartmann D., Astrophys. J. {\bf 812}, 142 (2016) \\ 
\medskip
Sitnova N.V., Mashonkina L.I., Astron. Lett. {\bf 44}, 411 (2018) \\
\medskip
Thielemann K.-F.,Argast D., Brachwitz F. et al., Astrophys. Space Sci. {\bf 281}, 25 (2002)\\
Tutukov A.V., Yungelson L.R., Mon. Not. R. Astr. Soc. {\bf 268}, 871 (1994)  \\
\medskip
Watanabe K., Hartman D.H., Leising M.D., The L.S.,
Astrophys. J. {\bf 516}, 285 (1999)   \\
\medskip
Watanabe K., Hartmann D.H., Leising M.D., The L.S., Share G.H., Kinzer R.L., 
Fourth Compton symposium. AIP Conference Proceedings. {\bf 410}, 1223 (1997)\\
\medskip  
Woosley S.E., Weaver T.A., Astrophys. J. Suppl., {\bf 101}, 181 (1995) \\
\medskip
Yungelson L.R.,  Astron. Lett. {\bf 36}, 780 (2010) \\
\medskip 
 Zhao, G., Mashonkina, L., Yan, H. L.  et al.,  Astrophys. J. {\bf 833}, 225 (2016)

\end{document}